\documentclass[preprint,showpacs,preprintnumbers,amsmath,amssymb,superscriptaddress]{revtex4}

\usepackage{graphicx}
\usepackage{dcolumn}
\usepackage{bm}
\usepackage{graphicx}
\usepackage{xspace}
\usepackage{natbib}
\usepackage{color}

\newcommand{\BKFAOP}{Ba$_{0.6}$K$_{0.4}$Fe$_2$As$_2$\xspace}
\newcommand{\TC}{\emph{T$_c$}\xspace}

\newcommand{\KF}{\emph{k}$_F$\xspace}

\newcommand{\KZ}{\emph{k}$_z$\xspace}
\newcommand{\KX}{\emph{k}$_x$\xspace}
\newcommand{\KY}{\emph{k}$_y$\xspace}

\begin{document}

\preprint{}

\title{Observation of a ubiquitous three-dimensional superconducting gap function in optimally-doped \BKFAOP}

\author{Y.-M. Xu}
\affiliation{Department of Physics, Boston College, Chestnut Hill, MA 02467, USA}
\author{Y.-B. Huang}
\affiliation{Beijing National Laboratory for Condensed Matter Physics, and Institute of Physics, Chinese Academy of Sciences, Beijing 100190, China}
\affiliation{Swiss Light Source, Paul Scherrer Institute, CH-5232 Villigen, Switzerland}
\author{X.-Y. Cui}
\affiliation{Swiss Light Source, Paul Scherrer Institute, CH-5232 Villigen, Switzerland}
\author{E. Razzoli}
\affiliation{Swiss Light Source, Paul Scherrer Institute, CH-5232 Villigen, Switzerland}
\author{M. Radovic}
\affiliation{Swiss Light Source, Paul Scherrer Institute, CH-5232 Villigen, Switzerland}
\author{M. Shi}
\affiliation{Swiss Light Source, Paul Scherrer Institute, CH-5232 Villigen, Switzerland}
\author{G.-F. Chen}
\affiliation{Department of Physics, Renmin University of China, Beijing 100872, China}
\author{P. Zheng}
\affiliation{Beijing National Laboratory for Condensed Matter Physics, and Institute of Physics, Chinese Academy of Sciences, Beijing 100190, China}
\author{N.-L. Wang}
\affiliation{Beijing National Laboratory for Condensed Matter Physics, and Institute of Physics, Chinese Academy of Sciences, Beijing 100190, China}
\author{C.-L. Zhang}
\affiliation{Department of Physics and Astronomy, The University of Tennessee, Knoxville, TN 37996, USA}
\author{P.-C. Dai}
\affiliation{Beijing National Laboratory for Condensed Matter Physics, and Institute of Physics, Chinese Academy of Sciences, Beijing 100190, China}
\affiliation{Department of Physics and Astronomy, The University of Tennessee, Knoxville, TN 37996, USA}
\affiliation{Neutron Scattering Sciences Division, Oak Ridge National Laboratory, Oak Ridge, TN 37831, USA}
\author{J.-P. Hu}
\affiliation{Beijing National Laboratory for Condensed Matter Physics, and Institute of Physics, Chinese Academy of Sciences, Beijing 100190, China}
\affiliation{Department of Physics, Purdue University, West Lafayette, IN 47907, USA}
\author{Z. Wang}
\affiliation{Department of Physics, Boston College, Chestnut Hill, MA 02467, USA}
\author{H. Ding}\email{dingh@aphy.iphy.ac.cn}
\affiliation{Beijing National Laboratory for Condensed Matter Physics, and Institute of Physics, Chinese Academy of Sciences, Beijing 100190, China}
\date{\today}

\maketitle

{\bf The iron-pnictide superconductors have a layered structureformed by stacks of FeAs planes from which the superconductivity originates. Given the multiband and quasi three-dimensional \cite{3D_SC} (3D)  electronic structure of these high-temperature superconductors, knowledge of the quasi-3D superconducting (SC) gap is essential for understanding the superconducting mechanism. By using the \KZ-capability of angle-resolved photoemission, we completely determined the SC gap on all five Fermi surfaces (FSs) in three dimensions on \BKFAOP samples. We found a marked \KZ dispersion of the SC gap, which can derive only from interlayer pairing. Remarkably, the SC energy gaps can be described by a single 3D gap function with two energy scales characterizing the strengths of intralayer $\Delta_1$ and interlayer $\Delta_2$ pairing. The anisotropy ratio $\Delta_2/\Delta_1$, determined from the gap function, is close to the $c$-axis anisotropy ratio of the magnetic exchange coupling $J_c/J_{ab}$ in the parent compound \cite{NeutronParent}. The ubiquitous gap function for all the 3D FSs reveals that pairing is short-ranged and strongly constrain the possible pairing force in the pnictides. A suitable candidate could arise from short-range antiferromagnetic fluctuations.}

Angle-resolved photoemision spectroscopy (ARPES) has played an important role in revealing the electronic structure of the pnictides. These measurements have typically been carried out at a fixed incident photon energy ($h\nu$) and varying incident angles that map out the planar band dispersion as a function of $k_x$ and $k_y$. Thus far, four Fermi-surface (FS) sheets have been observed with two hole pockets centred around the $\Gamma$ (0, 0) point and two electron pockets around the M ($\pi$, 0) point in the unfolded two-dimensional (2D) Brillouin zone. Below the superconducting transition temperature \TC, nodeless superconducting (SC) gaps open everywhere on the FS sheets \cite{HongARPES, Zhou_gap, Hasan_gap, NakayamaFourth}, pointing to a pairing order parameter with an s-wave symmetry in the $a$-$b$ plane, in agreement with a number of theoretical results \cite{Mazin, Kuroki, Hu, Lee, Tesanovic}. However there are other experiments which have indicated possible nodes in the superconducting gap of some pnictides, either line nodes in the $a$-$b$ plane or nodes along the $c$-axis \cite{Fletcher, Nakai, Reid}. It is well known that upon tuning the incident photon energy $h\nu$, the allowed direct transitions will shift in energy and consequently in the momentum perpendicular to the $a$-$b$ plane (\KZ) which enables the determination of the electronic dispersion along the $c$-axis. In the free electron final state approximation, the conversion is given by $k_z=\sqrt{2m[(h\nu-\phi-E_B)\cos^2 \theta+V_0]}/\hbar$, where $V_0$ is an experimentally determined inner potential \cite{hufner}. Several earlier ARPES studies have used this \KZ-resolving capability to probe the 3D dispersion of the normal state electronic structure in BaFe$_2$As$_2$-based pnictides, and found a large \KZ band dispersion in the orthorhombic phase in the vicinity of the parent compound \cite{Mannella, Kaminski, Fujimori, Fink, Brouet}. The 3D nature of the superconducting gap, which is critically important to the understanding of pnictide superconductivity, has yet to be studied.

Fig.~1a shows the spectral intensity measured at 10 K at a photon energy of 46 eV which corresponds to the reduced \KZ = 0 (see discussion below), plotted as a function of the binding energy and in-plane momentum along the $\Gamma$-X direction. Two hole-like bands are clearly observed, corresponding to the $\alpha$ (inner) and $\beta$ (outer) bands respectively, in both the spectral intensity plot and the second derivative intensity plot (Fig.~1b). The energy distribution curves (EDCs) show that the $\alpha$ band disperses towards the Fermi level ($E_F$) and bends back (Fig.~1c) as a result of the opening of the superconducting gap of $\sim$12 meV, consistent with earlier ARPES experiments \cite{HongARPES, NakayamaFourth}. When the photon energy is tuned to 32 eV (\KZ= $\pi$), in addition to the $\alpha$ and $\beta$ bands observed around the zone centre (the Z point), a third hole-like band (Fig.~1d) emerges between the $\alpha$ and $\beta$ bands, while the band calculations \cite{LDA_Singh,LDA_Lu,LDA_Xu} have predicted the existence of the third hole-like band in the pnictides, mostly for the 1111 pnictides, there is no accurate prediction of the observed \KZ dispersion of this band for the 122 compound. We label this new band as the $\alpha^{\prime}$ band, whose full dispersion and Fermi surface will be described in a separate paper.

Comparing Fig.~1b and Fig.~1d quantitatively, it is clear that the $\alpha$ band dispersion undergoes a parallel shifts in binding energy under different photon energies, as shown in Fig.~2a. Moreover, this modulation exhibits a well-defined periodicity in the photon energy, indicating that the excess energy is carried by the coherent interlayer quasiparticle (QP) tunneling with well defined momentum \KZ. Converting photon energy into momentum \KZ with an inner potential of 15 eV \cite{Mannella}, we find that the periodicity in \KZ is remarkably close to the one expected from the lattice spacing between the adjacent Fe layers, i.e. $2\pi/c^{\prime}$ = 0.951 \AA$^{-1}$, with $c^{\prime}$ =$c$/2 = 6.6 \AA ~(due to bilayer FeAs in \BKFAOP)  \cite{OPStructure}. To determine the \KZ dispersion of the $\alpha$ band, we stay at a fixed in-plane momentum such that the binding energy varies in a region sufficiently away from the SC gap, and plot the QP dispersion (Fig.~2b) as a function of \KZ as shown in Fig.~2c. The dispersion can be described remarkably well by
\begin{eqnarray}
E_{3D}^\alpha(k_x,k_y,k_z)=E_{2D}^\alpha(k_x,k_y) + 2t_\perp{\cos}k_z
\end{eqnarray}
with an interlayer hopping amplitude $t_\perp\simeq2.3$ meV in the binding energy range around 20 meV (along the $\Gamma$-X direction). Here and thereafter for notational simplicity, $k_x,k_y,k_z$ carries the units of $1/a, 1/b, 1/c^\prime$, respectively. The \KZ dispersion results in the warping of the $\alpha$ FS sheet along \KZ. One important implication of the observed \KZ dispersion is that this band cannot be a surface state which would have no \KZ dispersion. To describe the underlying FS quantitatively, we extrapolate the fitted dispersion given in Eq.~(1) to the Fermi level and display the in-plane Fermi wave vector ($k_F$) along $\Gamma$-X for the $\alpha$, $\beta$ and the electron-like ($\gamma/\delta$) bands as a function of \KZ in Figs.~2d$-$f, respectively. The FS area variation (defined as $\delta$ in ($1 \pm  \delta$)$A_{FS}$) in the $a$-$b$ plane are $\sim$10\%, 4\% and 1\% for the $\alpha$, $\beta$ and $\gamma/\delta$  bands, respectively, with the same periodicity along \KZ.

Having established the \KZ-dispersion of the quasi-3D electronic structure, we turn to the \KZ dependence of the SC gaps on different FS sheets obtained using many photon energies. Fig.~3a shows a collection of the EDCs at the Fermi crossings of the $\alpha$ band for different photon energies $h\nu=30 - 60$ eV. Appreciable gap variations are clearly visible in the symmetrised EDCs in Fig.~3b. The extracted SC gap values  (defined as $\Delta_{3D}^{\alpha}$) at both left and right Fermi crossings are plotted in Fig.~3c as a function of the photon energy (left axis) and c-axis momentum \KZ (right axis). Remarkably, $\Delta_{3D}^{\alpha}$ shows rather large periodic variations from $\sim$9 meV to $\sim$13 meV, then back to $\sim$ 9 meV as \KZ moves from Z to $\Gamma$ and back to Z. Similar \KZ dependence of the smaller SC gap on the $\beta$ band is also observed, varying from $\sim$5 meV to $\sim$7 meV, as shown later in Fig.~4. However, the \KZ variation of the SC gap on the electron-like ($\gamma/\delta$) FS sheets is much smaller, as indicated in Figs.~3d$-$f where the Bogoliubov QP peak is shown to situate at a nearly constant energy of $\sim$11.5 meV when \KZ varies from $\Gamma$ to Z. We note that an appreciable \KZ-dependence of pairing strength associated with 3D band structure has been predicted in this material \cite{3D_Scalapino}, although the predicted certain in-plane anisotropy and gap nodes have not been observed in our experiments.

It is interesting to note that the observed gap values (Figs.~3c and f) and the FS warping (Figs.~2d and f) along \KZ are anti-correlated, that is, when the in-plane FS area at a fixed \KZ becomes larger (smaller) the SC gap becomes smaller (larger) on the same plane. At the first glance, one might conclude that the gap variation along \KZ originates predominantly from the tunneling induced FS warping,that is, the \KZ dependence of the in-plane Fermi vector $k_F$, as is implied by the simplest form of an in-plane $s\pm$ gap function $\Delta_{s\pm}=\Delta_0\cos k_x \cos k_y$. However, this turns out to be not the case. The $<$ 10\% change in the in-plane Fermi vector is too small to account for the large, near 40\% gap variations because of the small ``gap velocity" in the pnictides. This is clearly seen from Fig.~1 where the near doubling of the Fermi vector in going from the $\alpha$-FS to the $\beta$-FS only results in the gap change from 12 meV to 6 meV. To illustrate this point further, we plot in Fig.~4a the measured gap values as a function of $\vert\cos k_x \cos k_y\vert$ at the Fermi points for the $\alpha$, $\alpha^{\prime}$, $\beta$, and the electron ($\gamma$/$\delta$) bands. Notice that although the average gap value follows this 2D form, the marked deviations induced by the \KZ dispersion could indicate that pairing is not purely two-dimensional and there is an additional driving force, namely, the pairing between the layers that is predominantly responsible for the gap dispersion with \KZ.

For an anisotropic layered superconductor with interlayer coupling, we adopt a simple form of the gap function,
\begin{eqnarray}
\Delta_{3D}(k_x,k_y,k_z)=\Delta_{2D}(k_x,k_y)(1+\eta {\cos}k_z).
\end{eqnarray}
This is a direct generalization of the expression for BCS superconductors with an isotropic in-plane gap function \cite{BZ}. Since the FS warping is rather small, $\Delta_{2D}(k_x,k_y)$ is approximately a constant. We expect this expression to be a good approximation, where $\eta$ is a measure of the ratio of the interlayer to in-plane pairing strength. In Fig.~4b, we plot the measured $\Delta_{3D}$ on different FS sheets as a function of cos\KZ. To increase the accuracy, we only used the lower photon energy part of the gap dispersion and averaged over the left and right crossings. Figure 4b shows that Eq.~(2) fits the data rather well, with the anisotropy ratio $\eta \sim$ 0.17, 0.13 and -0.01 for the $\alpha$, $\beta$ and $\gamma$/$\delta$ bands, respectively (the gap values of the $\alpha^{\prime}$ band are only resolvable near \KZ = $\pi$). It is interesting to note that the values of $\eta$ for the $\alpha$ and $\beta$ bands are consistent with the ratio of the exchange coupling $J_c/J_{ab}$ in the \text{magnetically ordered} parent compound extracted from the spin wave dispersion measured by neutron scattering \cite{NeutronParent}, where $J_c\sim5$ meV is the interlayer coupling and $J_{ab}\sim30$ meV is the next nearest neighbour coupling in the Fe-plane. Note that $k$-averaged $J_c$ would have most contributions from the bands with appreciate \KZ dispersion, such as $\alpha$, $\alpha^{\prime}$ and $\beta$ hole-like bands.

 The observation of the cosine dependence of the SC gap on \KX, \KY and \KZ indicates that the gap function reflects the lattice symmetry and that the predominant pairing is short-ranged in real space. Under such an assumption and taking into account the lattice symmetry, the leading terms of a generalized $s$-wave gap function can be written as $\Delta_1${cos\KX}cos\KY + $\frac{\Delta_{2}}{2}$(cos\KX + cos\KY)cos\KZ + $\Delta_3$({cos\KX}cos\KY)cos\KZ.  It is possible that the gap parameters of $\Delta_1$,  $\Delta_2$ and $\Delta_3$ have some band/orbital dependence. For simplicity, we choose band/orbital independent parameters to fit the data. We found that $\Delta_3 \ll \Delta_2$ as a result of the vanishingly small $\eta$ of the $\gamma$/$\delta$ electron-like FS around ($\pi$, 0). Remarkably, the remaining terms $\Delta_1${cos\KX}cos\KY + $\frac{\Delta_{2}}{2}$(cos\KX + cos\KY)cos\KZ fit all the gap values on the different FS sheets quite well, with $\Delta_1$ = 12.3 meV, and $\Delta_2$ = 2.07 meV as shown in Fig.~4c. The ratio of $\Delta_2/\Delta_1$ is nearly the same as the ratio of $J_c/J_{ab}$. We have also checked that the SC gap on each observed FS sheet along different in-plane directions fits well to this 3D gap function.
  
  Our finding of a single three-dimensional superconducting gap function for all five different FS pockets indicates that there are only two dominate pairing energy scales, one in-plane and one out-of-plane, in pnictide superconductors. It points to a common origin for the pairing strengths on all the observed FS sheets, independent of their different density of states. Moreover, because this gap function can be obtained by decoupling the 3D next nearest neighbour antiferromagnetic exchange couplings within the pairing channel, our results are consistent with short-ranged antiferromagnetic fluctuations induced pairing in the iron-pnictide superconductor \cite{Hu,Lee,si}.

\bigskip
\textbf{Methods}

We have performed systematic photon energy dependent ARPES measurements in the superconducting state of the optimally hole-doped \BKFAOP superconductors (\TC = 37 K). High quality single crystals used in our study were grown by the flux method \cite{ChenGF2}. High-resolution ARPES measurements were conduced at the SIS beamline of the Swiss Light Source. The photon energy used in the experiments is between 20 eV and 110 eV with different circular and linear polarization. The energy resolution is 8 - 20 meV depending on the photon energy, and the momentum resolution is below 0.02 $\AA^{-1}$. Samples were cleaved $\emph{in situ}$ and measured at 10 K in a working vacuum better than 5$\times10^{-11}$ Torr. The mirror-like sample surface was found to be stable without obvious degradation during a typical measurement period of 24 hours. Many samples have been measured and reproducible results have been obtained in these samples.

\newpage
\bigskip
\bigskip
\bigskip

\textbf{Author Contributions}

Y.-M.X., Y.-B.H., X.-Y.C., E.R. and M.R. performed experiments, Y.-M.X. and Y.-B.H. analysed data, H.D., Y.-M.X., J.-P.H. and Z.W. designed experiments, Z.W., H.D., Y.-M.X. and J.-P.H. wrote the paper, G.-F.C., P.Z., N.-L.W., C.-L. Z. and P.-C.D. synthesised materials. All authors discussed the results and commented on the manuscript.

\bigskip
\bigskip
\bigskip

\textbf{Corresponding author}

Correspondence to: H. Ding (dingh@aphy.iphy.ac.cn)

\bigskip
\bigskip
\bigskip

\textbf{Acknowledgements}

We thank valuable discussions of X. Dai, B. A. Bernevig and Z. Fang. This work was supported by grants from Chinese Academy of Sciences, NSF, Ministry of Science and Technology of China, NSF, DOE of US, and Sino-Swiss Science and Technology Cooperation.

\newpage
\begin{figure}[htbp]
\begin{center}
\includegraphics[width=10cm]{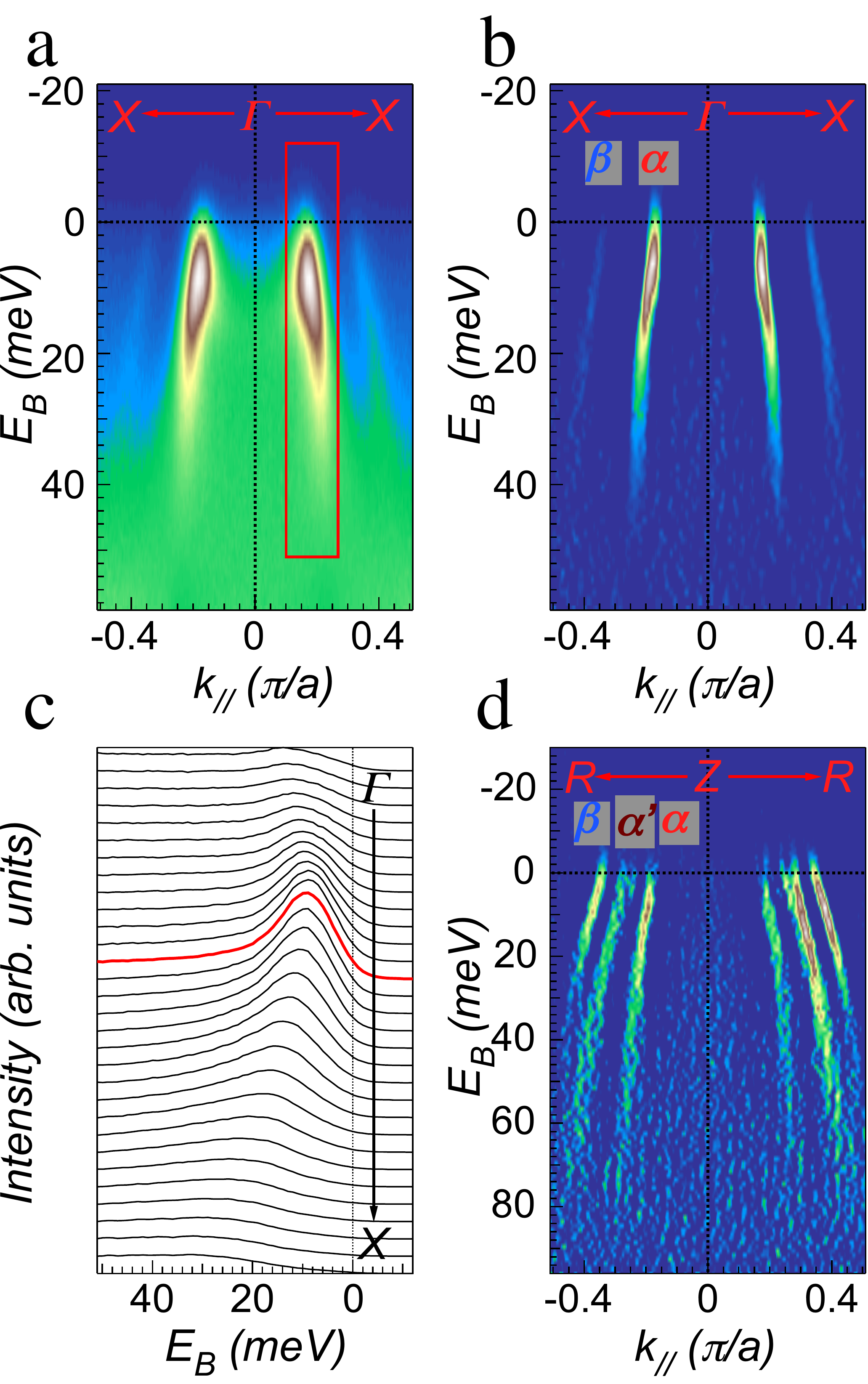}
\caption{\label{FIG01}  {\bf Band dispersion of superconducting \BKFAOP.}
$\bold{a,}$ ARPES spectral intensity measured at 10 K plotted in false color scale as a function of the in-plane momentum ($k_{\|}$) and binding energy along $\Gamma$-X by using 46-eV photons which corresponds to \KZ= 0. Two hole-like bands ($\alpha$ (inner) and $\beta$ (outer)) are observed. $\bold{b,}$ Second derivative of the spectral intensity plot as shown in panel $\bold{a}$. $\bold{c,}$ A set of EDCs within the $E$-$k$ range indicated by the red rectangle in panel $\bold{a}$. The red EDC is at \KF of the $\alpha$ band. $\bold{d,}$ Second derivative plot of the dispersion along Z-R (\KZ= $\pi$) measured by using 32-eV photons. Three hole-like bands ($\alpha$ (inner), $\alpha^{\prime}$ (middle) and $\beta$ (outer)) are observed.}
\end{center}
\end{figure}

\newpage
\begin{figure}[htbp]
\begin{center}
\includegraphics[width=12cm]{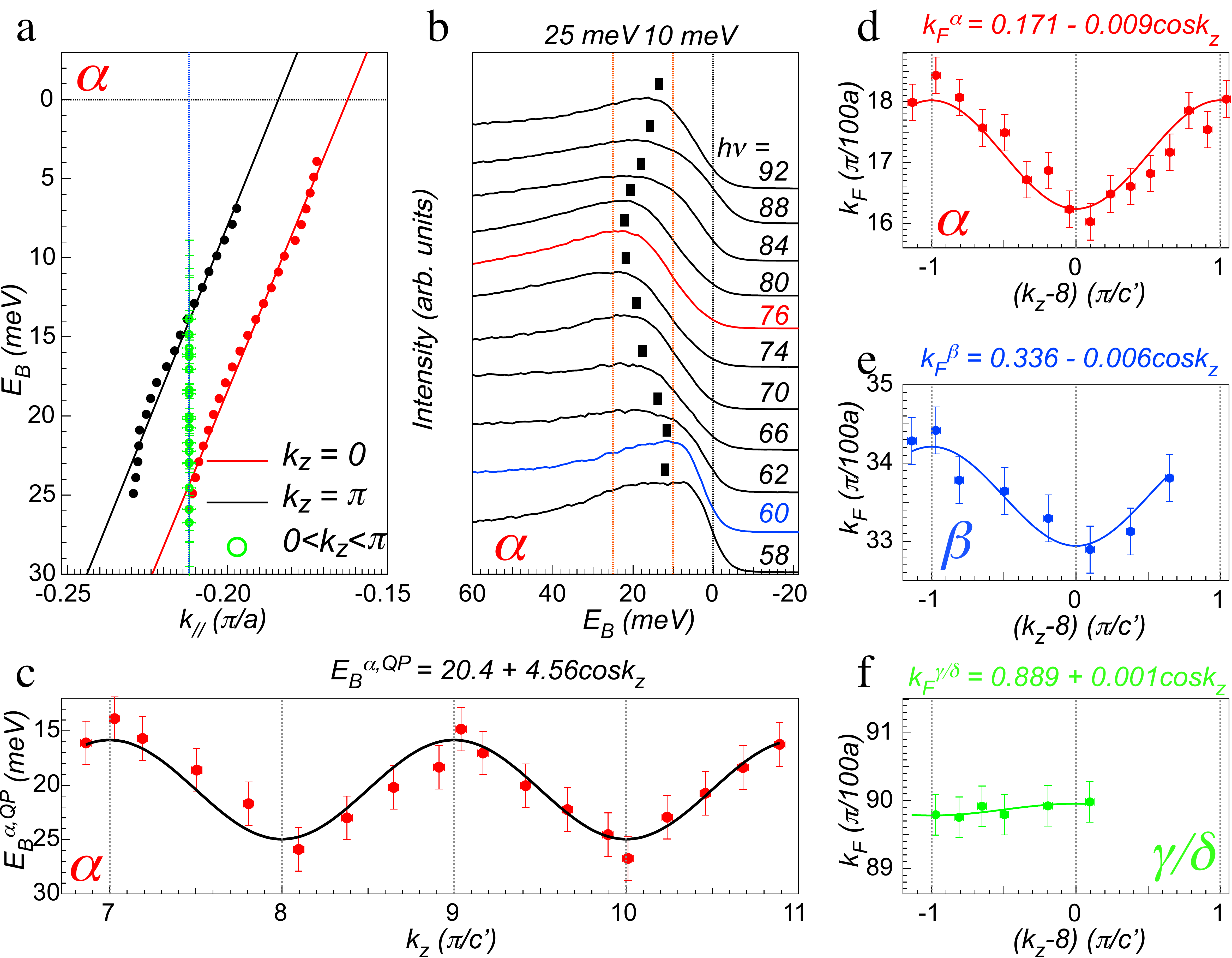}
\caption{\label{FIG02}  {\bf \KZ dispersion of quasiparticles and Fermi surface warping.}
$\bold{a,}$  Comparison of the $\alpha$ band dispersions at \KZ= 0 (red dots) and \KZ= $\pi$ (black dots) extracted from the peak position of momentum distribution curves along $\Gamma$-X. The solid lines are linear fits to the corresponding dispersions. The green dots denote the binding energies of the quasiparticle peak of the $\alpha$ band as shown in panel $\bold{b}$ measured at a fixed $k_{\|}$ (slightly below $k_F$) by using different photon energies (or different \KZ). $\bold{b,}$ Corresponding EDCs to the green dots shown in panel $\bold{a}$. The black vertical bars indicate the binding energies of the QP peaks. The red (blue) EDC corresponds to \KZ = 0 (\KZ = $\pi$).  Note that the ``shoulder" at the higher binding energy ($\sim$20 meV) of the EDC measured at $h\nu$ = 58 eV is from the $\alpha^{\prime}$ band. $\bold{c,}$ \KZ dispersion of the $\alpha$ band extracted from panels $\bold{a}$ and $\bold{b}$. The solid curve is the fit using cos\KZ. $\bold{d,}$ \KZ dependence of the in-plane \KF of the $\alpha$-FS parallel to $\Gamma$-X. The solid curve is the  cos\KZ fit. $\bold{e}$ and $\bold{f,}$ Same as panel $\bold{d}$, but for the $\beta$-FS and electron-like ($\gamma/\delta$) FS. }
\end{center}
\end{figure}

\newpage
\begin{figure}[htbp]
\begin{center}
\includegraphics[width=10cm]{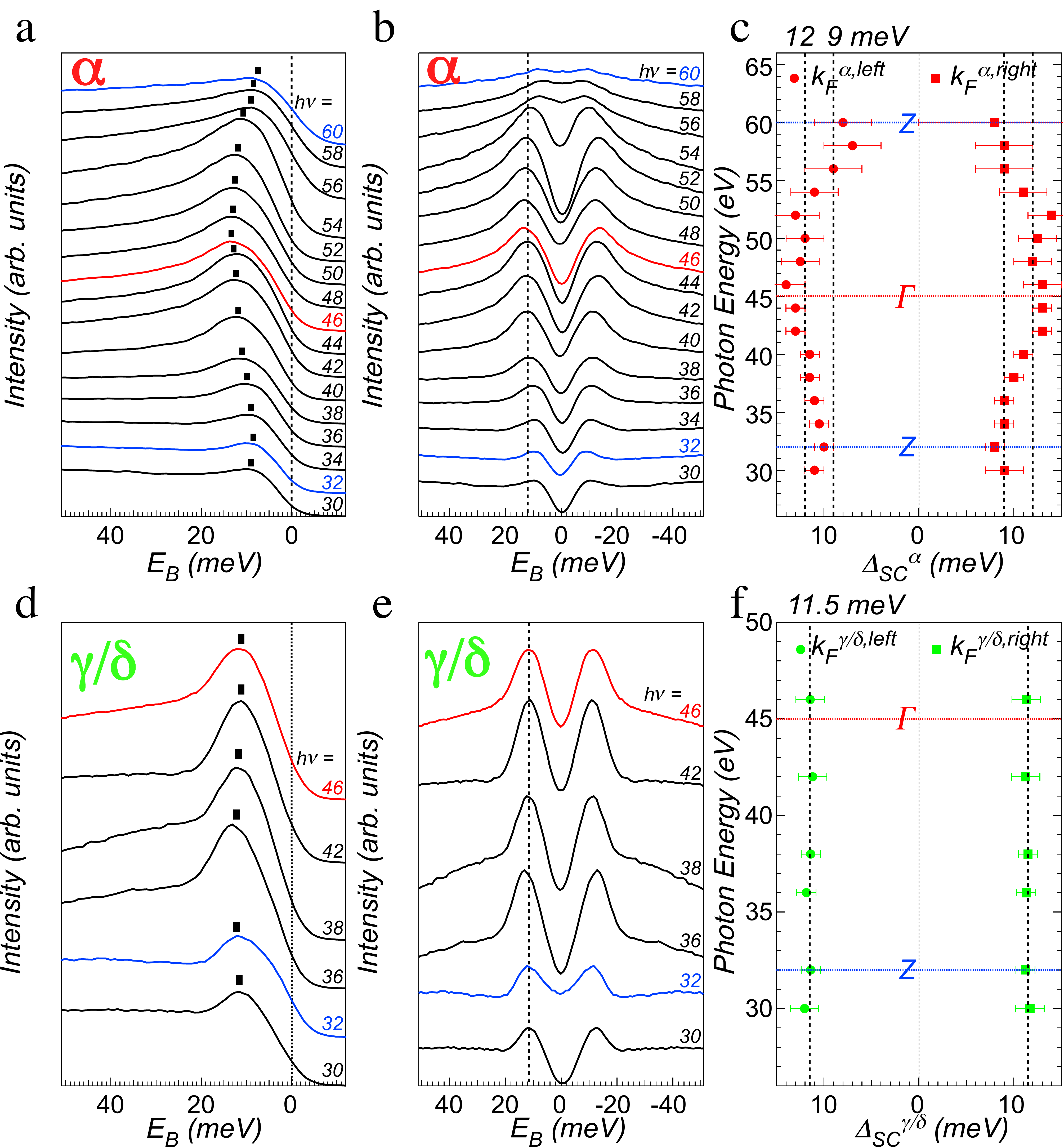}
\caption{\label{FIG03}   {\bf \KZ dependence of the superconducting gaps.}
$\bold{a,}$ Photon-energy-dependent EDCs measured at the left \KF on the $\alpha$-FS along $\Gamma$-X or its parallel directions with different \KZ. Red (blue) EDCs correspond to \KZ = 0 (\KZ = $\pi$).  The black blocks indicate the binding energies of the coherent peaks. $\bold{b,}$ Corresponding symmetrised EDCs of the ones shown in panel $\bold{a}$. The dashed line at 12 meV is a guide to eyes for viewing the variation of the SC gap at different $h\nu$. $\bold{c,}$ Extracted values of the SC gap (defined as the half value of peak-to-peak positions in symmetrised EDCs) on the $\alpha$-FS at different photon energies. The dots (squares) are obtained from left (right) side of \KF on the $\alpha$-FS. $\bold{d}$-$\bold{f,}$ Same as panels $\bold{a}$-$\bold{c}$ but on the electron-like FSs ($\gamma/\delta$) along $\Gamma$-M or its parallel directions. The dashed line in $\bold{e}$ is at 11.5 meV to guide eyes for viewing the SC gap at different $h\nu$. }
\end{center}
\end{figure}

\newpage
\begin{figure}[htbp]
\begin{center}
\includegraphics[width=15cm]{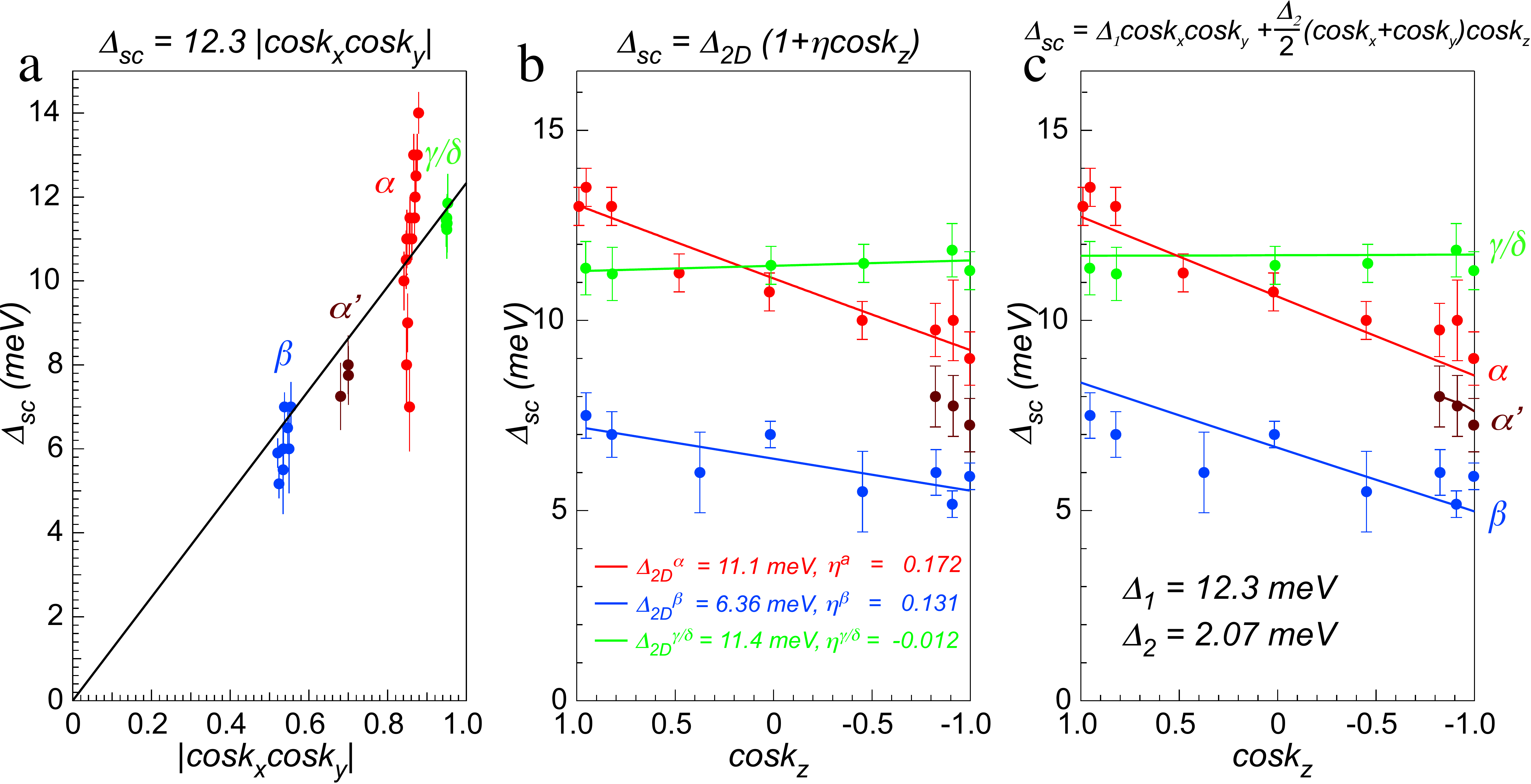}
\caption{\label{FIG04} {\bf 3D superconducting gap function $\Delta$(\KX, \KY, \KZ).}
$\bold{a,}$ The SC gap values on the $\alpha$-FS (red dots), $\beta$-FS (blue dots), $\gamma$/$\delta$-FS (green dots) and $\alpha^{\prime}$-FS (brown dots) as functions of $\vert\cos k_x \cos k_y\vert$. The black solid line is the gap function $|\Delta| = \Delta_{0}\vert\cos k_x \cos k_y\vert$ with $\Delta_{0}$ = 12.3 meV. $\bold{b,}$ Same as panel $\bold{a}$ but as functions of cos\KZ. The solid lines are independent linear fits to the SC gaps on the different FS sheets using a generic 3D gap function $|\Delta(k_x, k_y, k_z)|$ = $|\Delta_{2D}(k_x, k_y)$(1+$\eta$cos$k_z$)$|$. $\bold{c,}$ Same as panel $\bold{b}$ but using a single 3D gap function $|\Delta(k_x, k_y, k_z)| = |\Delta_{1}$cos{\KX}cos\KY + $\frac{\Delta_2}{2}$(cos\KX + cos\KY)cos$k_z|$ with $\Delta_1$  = 12.3 meV and $\Delta_2$ = 2.07 meV to fit all the SC gaps. }
\end{center}
\end{figure}

\end{document}